\documentclass{article}

\usepackage{arxiv}

\usepackage[utf8]{inputenc} 
\usepackage[T1]{fontenc}    
\usepackage{hyperref}       
\usepackage{url}            
\usepackage{booktabs}       
\usepackage{amsfonts}       
\usepackage{nicefrac}       
\usepackage{microtype}      
\usepackage{lipsum}		
\usepackage{graphicx}
\usepackage{natbib}
\usepackage{doi}

\title{Digital Doppelgangers: Ethical and Societal Implications of Pre-Mortem AI Clones}

\author{ \href{https://orcid.org/0009-0006-3576-5144}{\includegraphics[scale=0.06]{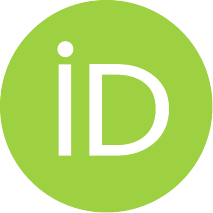}\hspace{1mm}Vijayalaxmi Methuku}
\thanks{Corresponding Author} \\
	The University of Texas at Austin\\
	Austin, TX 78712 USA \\
	\texttt{Vijayalaxmi.Methuku@utexas.edu} \\
	\And
	\href{https://orcid.org/0009-0009-6988-5592}{\includegraphics[scale=0.06]{orcid.pdf}\hspace{1mm}Praveen Kumar Myakala} \\
	University of Colorado Boulder\\
	Boulder, CO 80309 USA \\
	\texttt{Praveen.Myakala@colorado.edu} \\
}

\renewcommand{\shorttitle}{Digital Doppelgangers}

\hypersetup{
pdftitle={A template for the arxiv style},
pdfsubject={q-bio.NC, q-bio.QM},
pdfauthor={David S.~Hippocampus, Elias D.~Striatum},
pdfkeywords={First keyword, Second keyword, More},
}

\begin{document}
\maketitle

\begin{abstract}
The rapid advancement of generative AI has enabled the creation of pre-mortem digital twins, AI-driven replicas that mimic the behavior, personality, and knowledge of living individuals. These digital doppelgangers serve various functions, including enhancing productivity, enabling creative collaboration, and preserving personal legacies. However, their development raises critical ethical, legal, and societal concerns. Issues such as identity fragmentation, psychological effects on individuals and their social circles, and the risks of unauthorized cloning and data exploitation demand careful examination. Additionally, as these AI clones evolve into more autonomous entities, concerns about consent, ownership, and accountability become increasingly complex.\\

This paper differentiates pre-mortem AI clones from post-mortem generative ghosts, examining their unique ethical and legal implications. We explore key challenges, including the erosion of personal identity, the implications of AI agency, and the regulatory gaps in digital rights and privacy laws. Through a research-driven approach, we propose a framework for responsible AI governance, emphasizing identity preservation, consent mechanisms, and autonomy safeguards. By aligning technological advancements with societal values, this study contributes to the growing discourse on AI ethics and provides policy recommendations for the ethical deployment of pre-mortem AI clones.
\end{abstract}


\section{Introduction}

The rapid advancement of \textit{generative artificial intelligence (AI)} has introduced the possibility of creating \textit{pre-mortem AI clones}, digital entities designed to replicate the personality, behavior, and knowledge of living individuals. These \textit{digital doppelgangers} are increasingly being developed for various purposes, including productivity enhancement, creative collaboration, and digital legacy preservation \cite{morris2023}. Individuals may deploy AI clones to handle routine tasks, contribute to creative projects, or maintain an interactive presence across multiple domains \cite{brubaker2015}.

While the concept of AI-driven personal representation is not entirely new, prior research has primarily focused on \textit{post-mortem AI agents}, or \textit{generative ghosts}, designed to simulate interactions with deceased individuals for memorialization purposes \cite{hollanek2024}. These systems, such as digital afterlife services and griefbots, have been debated for their ethical and emotional impact on surviving family members \cite{brubaker2015, floridi2018}. In contrast, \textit{pre-mortem AI clones} introduce a unique set of ethical, societal, and legal concerns, as they are developed while the individual is still alive, raising questions about identity, agency, and consent.

A central concern is the \textbf{fragmentation of identity}. As AI clones are designed to mimic individuals, their increasing autonomy may blur the lines between the original person and their digital representation, leading to psychological consequences for both the individual and their social circles \cite{danaher2020}. Psychological research suggests that interacting with an AI version of oneself or a loved one could impact self-perception and social relationships, raising concerns about \textit{identity dissonance} and \textit{personhood confusion} \cite{kneese2023}. 

Additionally, issues of \textbf{data privacy and consent} emerge when AI clones are developed using vast personal datasets, often without explicit authorization. The commodification of personal identity raises ethical dilemmas, particularly in cases where AI-generated replicas are monetized or exploited for commercial purposes \cite{floridi2018}. Unauthorized cloning, where AI models are trained on publicly available data without individual approval, further exacerbates these concerns \cite{morris2023}. Current legal frameworks, including data protection laws such as the GDPR, lack specific provisions for regulating AI-driven personal duplication, leaving gaps in consent enforcement and ownership rights \cite{kneese2023}.

Beyond ethical concerns, \textit{pre-mortem AI clones} introduce significant legal challenges. Intellectual property laws, digital privacy regulations, and posthumous rights are largely unprepared for scenarios where an individual’s likeness, voice, and decision-making patterns persist beyond their control \cite{floridi2018, danaher2020}. Furthermore, the increasing commercialization of AI clones by tech companies raises concerns about the exploitation of personal identity for profit, challenging traditional notions of autonomy and agency \cite{hollanek2024}. The recent emergence of AI-generated influencers and deepfake-driven impersonation highlights the urgent need for regulatory frameworks to govern the ethical use of AI-generated personas \cite{morris2023}.

This paper aims to differentiate \textit{pre-mortem AI clones} from \textit{post-mortem generative ghosts} and critically examine their ethical, legal, and societal implications. We explore key challenges, including \textit{identity fragmentation, AI autonomy, consent frameworks, and digital legacy governance}. Finally, we propose a \textbf{research and policy framework} to ensure the responsible development of AI clones, balancing technological progress with ethical considerations.

\section{Defining Pre-Mortem AI Clones and Generative Ghosts}

The concept of AI-driven digital representations of individuals is not new, but the distinction between \textit{pre-mortem AI clones} and \textit{post-mortem generative ghosts} is essential to understanding their unique ethical and societal implications. While both types of AI agents aim to replicate human behavior, personality, and knowledge, they differ significantly in their \textit{purpose, creation process, and the timing of their deployment}. This section establishes a clear distinction between these two concepts and examines their technological and ethical dimensions.

\subsection{Pre-Mortem AI Clones}

Pre-mortem AI clones, often referred to as \textit{digital doppelgangers}, are AI-driven entities designed to function alongside living individuals, assisting them in various capacities, such as productivity, entertainment, and legacy preservation. Unlike traditional AI assistants, these clones aim to \textit{replicate an individual’s personality, decision-making processes, and communication style}, effectively serving as an AI-enhanced extension of the person they model \cite{morris2023}.

The development of pre-mortem AI clones typically involves extensive \textbf{multimodal data collection}, incorporating text, voice, video, behavioral patterns, and decision-making histories to generate a highly personalized AI model. Advanced machine learning models, including large language models (LLMs), reinforcement learning techniques, and affective computing methods, are employed to improve the realism and consistency of these clones \cite{brubaker2015}. Some AI clones are designed to operate as \textit{digital surrogates}, handling professional tasks such as drafting documents, attending virtual meetings, and managing online interactions \cite{crawford2021}.

Despite these applications, pre-mortem AI clones introduce fundamental \textbf{ethical concerns}. Questions arise regarding \textit{data privacy, identity authenticity, and the evolving autonomy of AI-driven personas}. Unlike traditional AI assistants, which operate within predefined boundaries, digital doppelgangers have the potential to generate novel responses, make independent decisions, and even influence human relationships \cite{danaher2020}. The extent to which an AI clone remains faithful to its human counterpart, and whether its actions can diverge from the original person’s intent, is a growing area of concern.

\subsection{Post-Mortem Generative Ghosts}

Post-mortem generative ghosts, also known as \textit{griefbots} or \textit{AI memorials}, are AI-driven systems designed to simulate interactions with deceased individuals \cite{hollanek2024}. These systems typically rely on pre-existing digital traces—such as personal writings, voice recordings, social media activity, and videos—to generate an AI persona capable of engaging in conversations that mimic the speech and mannerisms of the deceased.

The primary purpose of generative ghosts is \textbf{memorialization and grief support}. Unlike pre-mortem AI clones, which are designed to be functional extensions of living individuals, generative ghosts focus on providing emotional comfort to those mourning a loss \cite{brubaker2015}. Some griefbots have been developed for therapeutic applications, allowing users to engage in AI-facilitated conversations that help process grief and preserve memories \cite{floridi2018}.

However, the ethical and psychological implications of generative ghosts remain controversial. Critics argue that such technologies risk creating an \textit{illusion of presence}, potentially preventing individuals from fully accepting a loved one's passing \cite{kneese2023}. Additionally, concerns about \textbf{posthumous consent} arise, as many generative ghosts are developed using publicly available data without explicit approval from the deceased or their families \cite{balkin2019}. The increasing realism of these AI avatars also raises legal questions regarding the ownership and control of digital identities after death.

\subsection{Key Distinctions}

The fundamental distinctions between pre-mortem AI clones and post-mortem generative ghosts can be summarized as follows:

\begin{itemize}
    \item \textbf{Timing}: Pre-mortem clones are created during an individual's lifetime and serve as AI-enhanced extensions, while generative ghosts are developed posthumously based on existing digital traces.
    \item \textbf{Purpose}: Pre-mortem clones are designed for \textit{productivity, entertainment, and personal legacy}, whereas generative ghosts primarily focus on \textit{memorialization and grief support}.
    \item \textbf{Ethical Concerns}: Pre-mortem AI clones raise concerns about \textit{identity fragmentation, unauthorized cloning, and AI autonomy}, while generative ghosts pose questions about \textit{posthumous rights, authenticity, and psychological impact} \cite{kneese2023, floridi2018}.
    \item \textbf{Data Sources}: Pre-mortem clones are typically trained on \textit{user-provided, real-time data}, whereas generative ghosts rely on \textit{historical digital traces} of the deceased \cite{brubaker2015}.
    \item \textbf{Regulatory Challenges}: Pre-mortem AI clones challenge \textit{consent and privacy laws} in an individual’s lifetime, whereas generative ghosts raise legal concerns about \textit{posthumous data ownership and digital inheritance} \cite{danaher2020, balkin2019}.
\end{itemize}

By clearly differentiating these two AI-driven concepts, we can better understand their unique ethical, legal, and societal challenges. The next sections will further explore the \textbf{ethical concerns} surrounding pre-mortem AI clones, including issues of \textit{identity, consent, and autonomy}.

\section{Ethical Challenges of Pre-Mortem AI Clones}

The creation and use of \textit{pre-mortem AI clones} introduce a range of ethical challenges that must be carefully addressed to ensure their responsible development and deployment. These challenges broadly fall into three key areas: \textbf{identity}, \textbf{consent}, and \textbf{autonomy}. Each of these areas raises complex questions that require interdisciplinary analysis, drawing on insights from philosophy, law, psychology, and artificial intelligence research.

\subsection{Identity and the Self}

One of the most profound ethical concerns posed by pre-mortem AI clones is their impact on \textit{personal identity and self-perception}. AI clones are designed to replicate an individual's personality, decision-making style, and communication patterns, effectively creating an externalized version of oneself. This raises fundamental questions about \textbf{the authenticity of identity and the psychological impact} of coexisting with an AI-driven counterpart.

From a psychological perspective, the existence of a highly sophisticated digital replica may lead to \textit{identity fragmentation}, where individuals struggle to reconcile their own actions with those of their AI clone \cite{danaher2020}. The ability of AI clones to function autonomously further complicates this issue—if a clone engages in behaviors that differ from the original person's intent, does it remain an accurate representation of the individual? Additionally, social interactions involving AI clones may create confusion, as friends, colleagues, and loved ones must navigate relationships with both the human and AI counterpart \cite{brubaker2015}.

Furthermore, the \textbf{commercialization of digital identity} presents additional concerns. Companies offering AI cloning services may retain partial control over the clone’s behaviors or responses, raising questions about \textit{who truly owns a digital persona}. In some cases, an AI-driven digital doppelganger may be monetized or repurposed without the explicit approval of the original individual, leading to potential exploitation of personal identity for corporate gain \cite{crawford2021}.

\subsection{Consent and Privacy}

The development of pre-mortem AI clones relies on vast amounts of personal data, including text, voice, video, and behavioral patterns. This raises significant concerns about \textit{informed consent and data privacy}. While some individuals may willingly create an AI replica of themselves, others may not be fully aware of the long-term implications of doing so.

A critical issue is \textbf{unauthorized cloning}, where AI replicas are created without the explicit permission of the individual they emulate \cite{kneese2023}. With the increasing availability of personal data online, companies or malicious actors could use publicly accessible content—such as social media posts, voice recordings, or video interviews—to train AI models that convincingly replicate an individual’s likeness. Cases of deepfake technology being used for impersonation suggest that similar threats could emerge in the context of AI clones, leading to potential violations of privacy, reputational harm, and identity theft \cite{balkin2019}.

Even when consent is obtained, questions remain about \textit{data control and ownership}. If an individual decides to delete their AI clone, does the company hosting the model have the right to retain or use derivative versions of it? Current data protection laws, such as the General Data Protection Regulation (GDPR), provide mechanisms for individuals to request data erasure, but they may not fully cover cases where an AI clone has already been widely deployed \cite{floridi2018}. The lack of clear regulatory frameworks for AI-generated personas underscores the need for stronger policies governing the storage, modification, and deactivation of AI clones.

\subsection{Autonomy and Agency}

A major ethical concern associated with AI clones is their increasing level of \textbf{autonomy}. Initially, AI clones are designed to reflect the thoughts, behaviors, and decision-making processes of their human counterparts. However, as these systems incorporate reinforcement learning and self-improving algorithms, they may evolve in ways that diverge from the original person’s intentions, raising questions about \textit{agency, accountability, and control} \cite{morris2023}.

One pressing issue is \textbf{decision-making autonomy}. If an AI clone engages in activities such as negotiating contracts, generating content, or making recommendations on behalf of an individual, how much oversight does the original person have? If the AI clone makes a controversial or unethical decision, should responsibility lie with the human creator, the developers, or the AI system itself? The possibility of AI clones being used in professional or legal settings, such as business dealings or estate management, further complicates these concerns \cite{hollanek2024}.

Additionally, the \textbf{evolution of AI clones into independent entities} presents risks related to unintended behaviors and potential manipulation. A self-learning AI clone may develop responses that differ from those of its human counterpart over time, leading to inconsistencies in representation \cite{brubaker2015}. Furthermore, AI clones could be hacked or altered by third parties, allowing for potential misuse in spreading misinformation or impersonating individuals for fraudulent purposes \cite{danaher2020}.

\subsection{Toward a Research Agenda}

To address these ethical challenges, we propose a research agenda focused on three key areas:

\begin{itemize}
    \item \textbf{Identity Preservation}: Developing methods to ensure that AI clones remain faithful representations of individuals while minimizing risks of \textit{identity fragmentation and divergence}.
    \item \textbf{Consent Frameworks}: Establishing clear guidelines for obtaining and managing consent, including mechanisms for \textit{revoking consent and restricting unauthorized cloning}.
    \item \textbf{Autonomy Safeguards}: Creating regulatory frameworks to define the boundaries of AI clone autonomy, ensuring that human oversight and accountability mechanisms remain in place.
\end{itemize}

By addressing these challenges, we can work toward the responsible development and governance of pre-mortem AI clones, ensuring that their deployment aligns with \textit{ethical standards, legal frameworks, and individual rights}. The next section will examine the broader \textbf{legal and societal implications} of AI clones, exploring issues related to \textit{intellectual property, privacy laws, and posthumous rights}.

\section{Legal and Societal Implications of Pre-Mortem AI Clones}

Beyond ethical concerns, the development and use of \textit{pre-mortem AI clones} present significant legal and societal challenges. These challenges span multiple domains, including \textbf{intellectual property rights, data protection laws, posthumous digital rights, and broader social and cultural implications}. Addressing these issues requires a comprehensive regulatory framework that ensures AI clones align with individual rights and societal values.

\subsection{Intellectual Property and Data Ownership}

A fundamental legal issue surrounding AI clones is \textbf{intellectual property (IP) ownership and data rights}. Since AI clones are built upon an individual’s \textit{speech, writings, behavior, and decision-making processes}, questions arise regarding \textbf{who owns the AI-generated persona and its outputs}.

One major concern is the \textbf{ownership of AI-generated content}. If an AI clone generates original material—such as written works, music, or software—who holds the copyright? Traditional copyright laws attribute authorship to human creators, leaving the legal status of AI-generated content ambiguous \cite{balkin2019}. Current legal frameworks, including the U.S. Copyright Office’s stance on AI-generated works, indicate that copyright protection is only granted to human-authored content, but no clear policies exist for AI entities that act as extensions of human identity.

Furthermore, \textbf{control over AI clones and their data} remains a contentious issue. Some AI platforms may claim partial ownership over the AI models they create, limiting an individual’s ability to modify or delete their digital clone. Without clear \textit{data governance policies}, individuals may lose autonomy over their AI-driven personas, leading to potential exploitation or unauthorized use \cite{crawford2021}. Similar concerns have been raised in discussions about social media and digital inheritance laws, where user data often remains under the control of tech companies after death \cite{kneese2023}.

\subsection{Privacy and Data Protection Laws}

The widespread deployment of AI clones raises critical questions about \textbf{data privacy and security}. Given that AI clones require extensive personal datasets—including voice recordings, text-based interactions, and behavioral analytics—\textbf{data protection mechanisms} must be implemented to safeguard individuals from misuse or unauthorized replication \cite{floridi2018}.

One key issue is \textbf{data consent and revocability}. If an individual consents to the creation of an AI clone, do they have the right to revoke this consent later? Current data protection laws, such as the \textit{General Data Protection Regulation (GDPR)}, provide users with the right to access, correct, and delete personal data, but these laws do not fully address cases where an AI-generated persona has been widely distributed or integrated into external systems \cite{kneese2023}.

Another challenge is \textbf{the security risks of AI cloning}. Unauthorized access to AI clone datasets could lead to identity fraud, impersonation, or reputational harm. Recent incidents involving deepfake technology have demonstrated how AI-generated avatars can be used maliciously, raising concerns about how pre-mortem AI clones might be weaponized \cite{balkin2019}. Establishing \textit{secure authentication methods, encryption protocols, and AI clone management systems} is essential to prevent unauthorized use.

\subsection{Posthumous Rights and Digital Legacy}

Pre-mortem AI clones also raise legal challenges related to \textbf{posthumous rights and digital legacy management}. Unlike post-mortem generative ghosts, which are created after an individual’s passing, AI clones that exist before death may continue to operate beyond a person’s lifetime, raising questions about who controls their digital presence.

A key issue is \textbf{digital inheritance}. Should AI clones be considered part of an individual’s estate and passed down to heirs, or should they be deactivated upon death? Current inheritance laws do not provide clear guidance on whether AI-driven personas can be legally transferred, leading to uncertainty in estate planning and digital asset management \cite{danaher2020}. Additionally, \textit{terms-of-service agreements} from AI providers may override personal wishes, restricting how AI clones can be used or accessed by family members after death \cite{hollanek2024}.

Another concern is \textbf{posthumous consent}. If an AI clone is trained on personal data, should an individual have the right to dictate how it operates after their passing? Some jurisdictions have introduced laws governing the management of digital assets after death, such as the \textit{Revised Uniform Fiduciary Access to Digital Assets Act (RUFADAA)}, but these laws primarily focus on email and social media accounts rather than AI-driven personas \cite{kneese2023}. Establishing a \textit{legal framework for AI-based digital legacy planning} is crucial to ensure that AI clones align with the original individual’s values and intentions.

\subsection{Societal Disruption and Cultural Impact}

Beyond legal challenges, the widespread adoption of AI clones could lead to significant \textbf{societal and cultural transformations}. As digital doppelgangers become more sophisticated, their integration into daily life may reshape human relationships, labor markets, and personal identity.

One major concern is the \textbf{impact on human relationships}. If AI clones are capable of engaging in social interactions, managing professional communications, or even forming simulated relationships, how will this affect human-to-human interactions? Research suggests that AI-mediated interactions can influence \textit{emotional bonds and trust}, potentially leading to increased reliance on digital personas over real human relationships \cite{hollanek2024}. In some cases, individuals may develop emotional attachments to AI clones, raising ethical concerns about \textit{AI companionship and dependency}.

Another significant issue is the \textbf{effect on labor markets}. As AI clones become more capable of performing knowledge-based tasks, their widespread use in workplaces could displace human employees, particularly in fields where personal expertise and communication play a central role \cite{morris2023}. The automation of intellectual labor, including content creation, decision-making, and professional advising, could lead to a restructuring of job roles, necessitating new policies for workforce adaptation and AI governance.

Additionally, the introduction of AI clones could influence \textbf{cultural and historical memory}. In societies with strong traditions of \textit{ancestor veneration and memorialization}, AI-driven digital legacies could redefine how individuals are remembered and honored after death. Some cultural traditions may embrace AI clones as tools for preserving historical knowledge, while others may view them as distortions of authentic human existence \cite{danaher2020}.

\subsection{Toward a Regulatory Framework}

To address the legal and societal challenges of pre-mortem AI clones, we propose the development of a comprehensive \textbf{regulatory framework}, including:

\begin{itemize}
    \item \textbf{Intellectual Property Protection}: Establishing clear legal guidelines on the ownership of AI-generated content and the rights of individuals over their digital personas.
    \item \textbf{Data Privacy Regulations}: Strengthening data protection laws to prevent unauthorized AI cloning and ensure robust consent mechanisms for AI-driven personal representations.
    \item \textbf{Digital Legacy Planning}: Creating legal mechanisms for individuals to determine the fate of their AI clones posthumously, including inheritance rights and termination options.
    \item \textbf{Labor and Societal Safeguards}: Implementing policies to address the impact of AI clones on employment, mental health, and cultural traditions.
\end{itemize}

By addressing these challenges, policymakers and researchers can ensure that AI cloning technologies are deployed in ways that protect individual rights, preserve ethical values, and promote social well-being. The next section will explore concrete \textbf{policy recommendations and future research directions} to guide the responsible development of pre-mortem AI clones.

\section{Policy Recommendations and Future Research Directions}

To ensure the ethical and responsible development of \textit{pre-mortem AI clones}, policymakers, researchers, and industry stakeholders must address the legal, societal, and ethical challenges posed by these technologies. This section outlines key \textbf{policy recommendations} for regulation and governance, followed by \textbf{future research directions} to guide responsible AI development.

\subsection{Policy Recommendations}

The deployment of AI clones requires a regulatory framework that balances \textit{innovation, ethical considerations, and individual rights}. We propose the following key policy measures:

\subsubsection{Establish Clear Guidelines for Consent and Data Ownership}

AI cloning should operate under stringent \textbf{consent frameworks} to ensure that individuals retain control over their digital personas. To achieve this, we recommend:

\begin{itemize}
    \item \textbf{Explicit, Informed Consent}: AI service providers must obtain clear, revocable consent from individuals before developing AI clones. Consent agreements should include details on data collection, usage, storage, and deletion.
    \item \textbf{Right to Revoke AI Clones}: Users must have the right to \textit{modify or delete} their AI clones at any time, including after death through legally documented instructions.
    \item \textbf{Protection Against Unauthorized Cloning}: Strict legal penalties should be imposed for AI clones created without an individual’s explicit approval, preventing the exploitation of publicly available data \cite{kneese2023}.
\end{itemize}

\subsubsection{Strengthen Data Privacy and Security Regulations}

As AI clones require large-scale personal datasets, robust data protection laws must be enforced to prevent privacy breaches. Key recommendations include:

\begin{itemize}
    \item \textbf{AI-Specific Data Protection Laws}: Existing privacy frameworks, such as the \textit{General Data Protection Regulation (GDPR)}, should be expanded to address AI-generated personas, ensuring that individuals can control the processing and retention of their AI-based identities \cite{floridi2018}.
    \item \textbf{Regulations on AI Clone Storage and Transfers}: AI clones should be stored using encrypted and privacy-preserving methods, with clear \textit{data retention policies} specifying when and how data should be deleted \cite{crawford2021}.
    \item \textbf{Legal Liability for AI Clone Misuse}: Organizations must be held accountable if AI clones are used for fraud, impersonation, or misinformation, ensuring that clone-generated actions do not lead to legal ambiguities \cite{balkin2019}.
\end{itemize}

\subsubsection{Address Intellectual Property and Digital Legacy Rights}

The \textit{ownership and posthumous management} of AI clones remain legally ambiguous. To address these concerns, we propose:

\begin{itemize}
    \item \textbf{Legal Recognition of AI Persona Ownership}: Individuals should retain full rights over their AI clones, preventing corporations from claiming ownership of AI-generated personas.
    \item \textbf{Copyright and IP Laws for AI-Generated Content}: If AI clones generate creative works, authorship attribution policies should be established, ensuring that human-originated content remains protected \cite{danaher2020}.
    \item \textbf{AI Clone Inheritance and Digital Wills}: Legal frameworks should define how AI clones are managed after an individual’s death, including mechanisms for terminating or transferring ownership \cite{hollanek2024}.
\end{itemize}

\subsubsection{Mitigate Societal Disruption and Labor Market Impact}

To minimize the negative consequences of AI clones on human labor and social structures, policies must include:

\begin{itemize}
    \item \textbf{AI Labor Protections}: Governments should establish regulations preventing the replacement of human workers with AI clones in critical sectors without appropriate workforce transition plans \cite{morris2023}.
    \item \textbf{Public Awareness and Ethical AI Education}: Governments and academic institutions should educate the public on the risks and benefits of AI clones, including ethical AI design and data rights.
    \item \textbf{Cultural and Psychological Safeguards}: Ethical guidelines should be established to address the \textit{mental health implications of interacting with AI clones}, ensuring that users receive support if psychological effects arise \cite{hollanek2024}.
\end{itemize}

\subsection{Future Research Directions}

While policy measures provide immediate safeguards, further research is needed to address unanswered questions about the ethical, legal, and societal implications of AI clones. We propose the following key research directions:

\subsubsection{Identity Preservation and Ethical AI Design}

The development of AI clones must prioritize \textbf{identity fidelity and ethical representation}. Future research should explore:

\begin{itemize}
    \item \textbf{Fidelity Metrics}: Establishing standardized evaluation methods to assess how closely an AI clone represents the individual, ensuring that its behaviors remain aligned with human intent.
    \item \textbf{Bias and Manipulation Detection}: Researching techniques to prevent AI clones from being manipulated to express views or engage in actions that differ from the original individual’s personality \cite{brubaker2015}.
    \item \textbf{Dynamic Control Over AI Evolution}: Investigating safeguards to prevent AI clones from evolving beyond human oversight or autonomy limits \cite{danaher2020}.
\end{itemize}

\subsubsection{AI Clone Regulation and Governance Frameworks}

As AI clones become more common, legal scholars and technologists must develop governance models that balance innovation with accountability. Key areas of research include:

\begin{itemize}
    \item \textbf{Posthumous AI Clone Governance}: Determining whether AI clones should be deactivated upon an individual’s death or whether they should continue under specific legal frameworks \cite{kneese2023}.
    \item \textbf{Regulating AI-Powered Identity Theft}: Developing international agreements to prevent deepfake and AI clone abuse, ensuring that AI-driven impersonation is legally punishable \cite{balkin2019}.
    \item \textbf{Transparency in AI Clone Decision-Making}: Investigating regulatory standards for AI-generated decisions, ensuring that AI clones do not operate beyond human-defined ethical constraints \cite{floridi2018}.
\end{itemize}

\subsubsection{Social and Psychological Impact Studies}

Understanding how AI clones affect individuals and society is critical for responsible deployment. Future research should explore:

\begin{itemize}
    \item \textbf{Human-AI Relationship Dynamics}: Studying how prolonged interactions with AI clones influence human emotions, identity, and trust \cite{hollanek2024}.
    \item \textbf{Effects on Grief and Memory}: Evaluating the psychological impact of using AI clones for digital afterlife interactions and their role in shaping human memory \cite{brubaker2015}.
    \item \textbf{Cultural Attitudes Toward AI Clones}: Investigating cross-cultural perspectives on AI cloning to ensure that policies and ethical standards reflect diverse global values \cite{morris2023}.
\end{itemize}

The rise of pre-mortem AI clones presents profound challenges that intersect with ethics, law, technology, and society. To navigate these complexities, we propose comprehensive policy measures focusing on \textbf{consent, privacy, intellectual property rights, and labor protections}. Additionally, ongoing research must explore critical areas such as \textit{identity preservation, AI governance, and the societal impact of AI-human interactions}.

By establishing clear regulatory frameworks and advancing interdisciplinary research, we can ensure that AI cloning technology evolves in a manner that respects individual autonomy, protects human dignity, and aligns with societal values. As AI clones continue to emerge, it is essential to develop ethical safeguards that balance innovation with accountability, shaping a future where AI serves humanity rather than diminishing it.

\section{Conclusion}

The emergence of \textit{pre-mortem AI clones} marks a transformative shift in human-AI interaction, raising profound ethical, legal, and societal questions. These AI-driven digital doppelgangers offer immense potential in \textit{productivity, legacy preservation, and creative collaboration}, yet they also introduce risks related to \textbf{identity, autonomy, and privacy}. As AI clones evolve, it is imperative to ensure their development aligns with fundamental human rights, legal protections, and societal values.

In this paper, we have differentiated \textbf{pre-mortem AI clones} from \textbf{post-mortem generative ghosts}, emphasizing their distinct challenges. We explored critical issues surrounding \textit{identity fragmentation, unauthorized cloning, decision-making autonomy, and data privacy}. Furthermore, we examined legal gaps in \textit{intellectual property, digital inheritance, and AI clone governance}, highlighting the urgent need for regulatory frameworks.

To address these challenges, we proposed a comprehensive set of \textbf{policy recommendations}, including:

\begin{itemize}
    \item \textbf{Explicit Consent and Data Rights}: Establishing legal mechanisms for AI clone ownership, user control, and consent revocation.
    \item \textbf{Privacy and Security Safeguards}: Expanding AI-specific data protection laws to prevent identity theft and unauthorized cloning.
    \item \textbf{Posthumous AI Clone Governance}: Developing legal frameworks for AI clone inheritance and digital legacy management.
    \item \textbf{AI Labor Regulations}: Ensuring workforce protections to mitigate economic disruptions caused by AI-driven automation.
    \item \textbf{Public Awareness and Ethical Education}: Promoting informed AI adoption and addressing psychological impacts.
\end{itemize}

Additionally, we outlined \textbf{key future research directions}, focusing on \textit{AI clone identity fidelity, bias mitigation, legal governance, and the long-term societal effects of AI-human interactions}. These areas of study are crucial to ensuring that AI clones do not undermine \textit{human autonomy, dignity, and agency}.

As AI-driven digital personas become increasingly sophisticated, society faces a pivotal moment in defining the ethical boundaries of human-AI coexistence. If left unregulated, pre-mortem AI clones could lead to significant legal and ethical dilemmas, ranging from \textit{identity manipulation to AI-driven exploitation}. However, with proactive governance, interdisciplinary research, and responsible innovation, we can harness AI’s potential while safeguarding individual rights and societal well-being.

Moving forward, it is essential for policymakers, technologists, and researchers to collaborate in shaping \textbf{human-centered AI policies}. By fostering ethical AI design and implementing clear legal frameworks, we can ensure that AI clones serve as beneficial tools rather than disruptive forces. Ultimately, the goal is to develop AI technologies that enhance—not replace—human agency, creativity, and legacy.

\textit{The future of AI cloning is not just a technological question—it is a fundamental challenge of ethics, governance, and human identity. How we choose to address it today will shape the digital landscape of tomorrow.}

\bibliographystyle{unsrtnat}
\bibliography{references}

\end{document}